\documentclass[conference]{IEEEtran}
\IEEEoverridecommandlockouts
\newtheorem{definition}{Definition}
\usepackage{cite}
\usepackage{amsmath,amssymb,amsfonts}
\usepackage{algorithmic}
\usepackage{fancyhdr}
\usepackage{algorithm}
\usepackage{subfigure}
\usepackage{booktabs}
\usepackage{color}
\usepackage{graphicx}
\usepackage{textcomp}
\usepackage{xcolor}
\usepackage{comment}
\def\BibTeX{{\rm B\kern-.05em{\sc i\kern-.025em b}\kern-.08em
    T\kern-.1667em\lower.7ex\hbox{E}\kern-.125emX}}
\newcommand{\equalcontribution}{\textsuperscript{*}}
\newcommand{\correspondingauthor}{\textsuperscript{$\dagger$}}
\renewcommand{\footnoterule}{%
  \kern -3pt
  \hrule width 0.49\textwidth height 0.4pt
  \kern 2.6pt
}

\begin{document}

\title{FedPDD: A Privacy-preserving Double Distillation Framework for Cross-silo Federated Recommendation
}
\author{
\IEEEauthorblockN{Sheng Wan\equalcontribution}
\IEEEauthorblockA{\textit{Dept. of CSE} \\
\textit{SUSTech \& HKUST}\\
Hong Kong, China \\
swanae@cse.ust.hk}
\and
\IEEEauthorblockN{Dashan Gao\equalcontribution
\protect\thanks{*These authors contributed equally to this work.}
}
\IEEEauthorblockA{\textit{Dept. of CSE} \\
\textit{SUSTech \& HKUST}\\
Hong Kong, China \\
dgaoaa@cse.ust.hk}
\and
\IEEEauthorblockN{Hanlin Gu}
\IEEEauthorblockA{\textit{Webank}\\
Shenzhen, China \\
ghltsl123@gmail.com}
\and
\IEEEauthorblockN{Daning Hu\correspondingauthor
\protect\thanks{\textsuperscript{$\dagger$}Corresponding author.}}
\IEEEauthorblockA{\textit{Dept. of Finance} \\
\textit{SUSTech }\\
Shenzhen, China \\
hdaning@gmail.com}
}

\maketitle

\begin{abstract}
Cross-platform recommendation aims to improve recommendation accuracy by gathering heterogeneous features from different platforms. However, such cross-silo collaborations between platforms are restricted by increasingly stringent privacy protection regulations, thus data cannot be aggregated for training. Federated learning (FL) is a practical solution to deal with the data silo problem in recommendation scenarios. Existing cross-silo FL methods transmit model information to collaboratively build a global model by leveraging the data of overlapped users. However, in reality, the number of overlapped users is often very small, thus largely limiting the performance of such approaches. Moreover, transmitting model information during training requires high communication costs and may cause serious privacy leakage. In this paper, we propose a novel privacy-preserving double distillation framework named FedPDD for cross-silo federated recommendation, which efficiently transfers knowledge when overlapped users are limited. Specifically, our double distillation strategy enables local models to learn not only explicit knowledge from the other party but also implicit knowledge from its past predictions. Moreover, to ensure privacy and high efficiency, we employ an offline training scheme to reduce communication needs and privacy leakage risk. In addition, we adopt differential privacy to further protect the transmitted information. The experiments on two real-world recommendation datasets, HetRec-MovieLens and Criteo, demonstrate the effectiveness of FedPDD compared to the state-of-the-art approaches.
\end{abstract}


\section{Introduction}
Benefiting from the explosion of data, deep learning-based recommendation systems have gained significant attention by overcoming obstacles of conventional models and achieving high recommendation quality \cite{zhang2019deep}. Unfortunately, in reality, this wealth of data is often separated into different platforms and owned by different entities. For example, people can chat with friends on WhatsApp, watch favorite videos on TikTok or Youtube, and buy wanted stuff on Amazon. Collecting these features from different platforms can help to build a more accurate user profile and provide a better recommendation. However, cross-silo collaborations among different platforms are restricted by data protection regulations such as General Data Protection Regulation (GDPR) and data cannot be centralized for training.

To tackle the privacy issue for cross-silo recommendation, a practical solution is Federated Learning (FL) \cite{yang2019federated,kairouz2019advances}. FL enables multiple parties to collaboratively train a global model while private data resides locally on the data owners and therefore can largely reduce systemic privacy risks. Existing cross-silo FL methods \cite{9076003,Siwei2020mmvfl,kang2020fedmvt} try to fix this problem by using the overlapped samples across participants and viewing it as a multi-view learning problem. The performance of such approaches highly relies on the number of overlapped users between parties. However, in reality, such overlapped data is often limited and thereby may cause the performance to be even worse than the locally trained models. Moreover, these approaches transmit feature or model information during training, which requires high communication costs and has serious privacy weaknesses. Recent studies \cite{zhu2019deep,geiping2020inverting,sun2021soteria} show that sharing model or feature information could still lead to private data leakage.

\begin{figure}[t]
  \centering
  \includegraphics[width=\columnwidth]{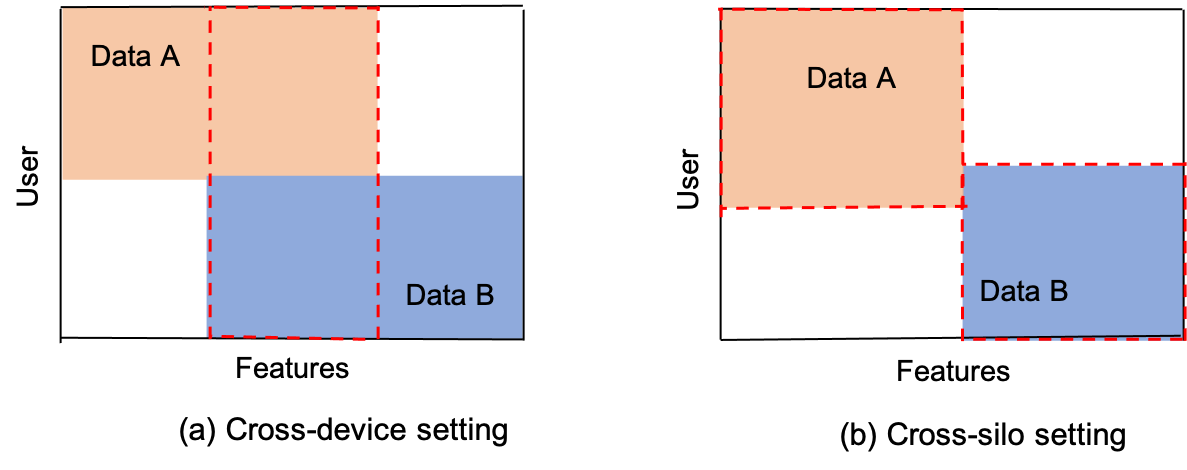}
  \vspace{-2em}
  \caption{In cross-device FL setting, participants are a large number of individual customers (2C) that have the same feature space. In cross-silo FL setting, participants are a small number of business partners (2B) that have partially overlapped user spaces and different feature spaces. Here we assume that there is no feature overlapping in our cross-silo FL setting. Data in the red box are used for training.}
  \label{settingpic}
  \vspace{-1em}
\end{figure}

To address these challenges, we propose a novel privacy-preserving double distillation framework named FedPDD for cross-silo federated recommendation. We design a double distillation strategy, which enables local models to learn both implicit knowledge from themselves and explicit knowledge from the other party. Specifically, we distill implicit knowledge from the past predictions of local models and distill explicit knowledge from the ensemble predictions of current local models. The key idea is that we provide multiple informative sources for local model training. Therefore, by learning from these sources, FedPDD is able to enhance model performance and generalization ability. Moreover, we employ an offline distillation strategy and only transmit model output during training. Parties only communicate with the server during the federated ensemble stage and the size of the model output is much smaller than the model itself. Accordingly, our training strategy largely reduces communication needs and limits the exposure of private information to the server. In addition, we adopt differential privacy \cite{abadi2016deep} to further protect the communication process. We experiment on two real-world recommendation datasets showing that FedPDD significantly boosts local model performance by up to 3.94\% and outperforms the state-of-the-arts by up to 3.98\%. 

Overall, our main contributions are as follows:
\begin{itemize}
    \item We propose a novel privacy-preserving double distillation method named FedPDD for cross-silo federated recommendation. Our method enables local models to learn from not only private labels and the other party but also itself, which enhances model performance and generalization ability when overlapped samples are limited.
    \item We employ an offline training strategy to reduce communication needs and privacy leakage risk. Moreover, we adopt differential privacy to further protect the communication process and provide a theoretical privacy analysis of FedPDD.
    \item We conduct experiments on two public real-world datasets and the results demonstrate the effectiveness of FedPDD by up to 3.98\% further improvements compared to the state-of-the-art approaches.
\end{itemize}

\section{Background and Related Work}\label{related}
\subsection{Federated Learning}
Federated Learning \cite{kairouz2019advances,yang2019federated} allows multiple participants to collaboratively train a global model while keeping training data in local. The data resides locally on the data owners and therefore largely reduces systemic privacy risks. Based on scenarios, FL can be divided into two kinds of settings: cross-device FL and cross-silo FL \cite{kairouz2019advances}. As shown in Figure 1, in the cross-device setting, participants are a large number of individual customers (2C) that have the same feature space, while participants are a small number of business partners (2B) that have partially overlapped user spaces and different feature spaces in the cross-silo setting. In this work, we focus on the cross-silo setting that features are not overlapped.

Liu et al. \cite{9076003} first proposed a transfer learning method named federated transfer learning (FTL) to transfer knowledge through the overlapped user data between parties. They assumed that only one party owns the labels and aims to improve the model performance by leveraging the knowledge (i.e., features) from other parties. Under such an assumption, their approach merely leveraged the overlapped users across parties. Existing cross-silo FL methods \cite{Siwei2020mmvfl,kang2020fedmvt} mostly follow this direction which views the distributed features of overlapped users in different parties as different views of these data and regards it as a multi-view learning problem. Specifically, Feng et al. \cite{Siwei2020mmvfl} established a Multi-participant Multi-class Vertical Federated Learning (MMVFL) framework. They utilized multi-view learning methods to securely share label information between participants. Kang et al. \cite{kang2020fedmvt} proposed a self-supervised multi-view learning method called FedMVT under the cross-silo FL setting. They built a model based on the overlapped data to predict the missing features. However, these studies build models on the overlapped user data across parties and prediction accuracy of the global model highly relies on the amount of overlapped data. When such data is limited, the performance of these approaches may be even worse than the local fine-tuned models. In contrast, FedPDD leverages both the overlapped data and the non-overlapped data to enhance model performance through knowledge distillation and can achieve superior performance compared to the state-of-the-arts when overlapped data is limited.

\begin{table}[t]\centering
  \caption{Definition of Notations}
  \label{tab1}
  \begin{tabular}{cc}
    \toprule
    Notation & Description\\
    \midrule
    $D^k$ & local dataset on party $k$\\
    $D^c$ & overlapped data shared between two parties\\
    $\alpha$ & ratio of overlapped data to the training set\\
    $f^k$ & local model on party $k$ \\
    $f^k_{b_{(n)}}$ & best local model on party $k$ in round $n$ \\
    $z^k$ & logit output of local model $f^k$ \\
    $z^t$ & ensemble logit \\
    $z'^t$ & perturbed ensemble logit \\
    $p^k$ & soft target output of local model $f^k$ \\
    $p^t$ & ensemble teacher knowledge \\
    $p^t_b$ & self teacher knowledge \\
    $\sigma_T$ & general softmax function \\
    $T$ & distillation temperature \\
  \bottomrule
\end{tabular}
\vspace{-2em}
\end{table}

\begin{figure*}[t]
  \centering
  \includegraphics[width=2\columnwidth]{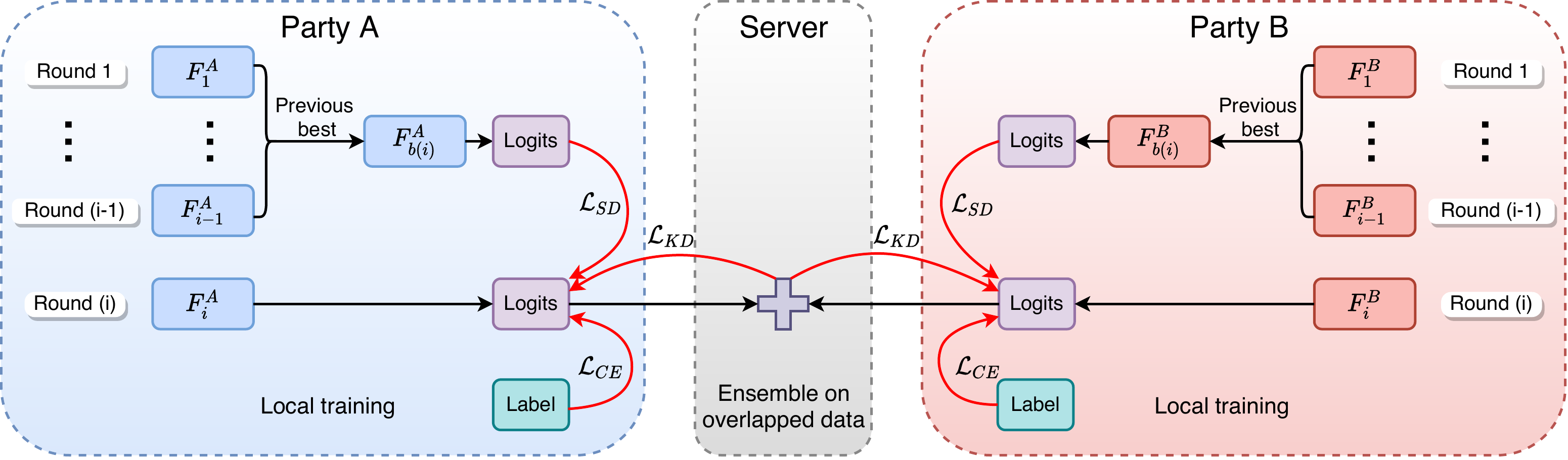}
  \caption{The overview of our proposed FedPDD. During training, each party trains its local model via three kinds of knowledge from the ground truth labels, ensemble of local models and past predictions of local models.  
  }
  \label{img_pipeline}
\vspace{-1em}
\end{figure*}

\vspace{-1mm}
\subsection{Federated Recommendation System}
Inspired by the success of federated learning \cite{mcmahan2017communication}, federated recommendation system is proposed to address the privacy and data silo problems in the recommendation system \cite{yang2019federated}. Existing federated recommendation works such as federated matrix factorization methods \cite{ying2020shared,yang2021practical,chai2020secure,yang2021fcmf,eren2022fedsplit} and federated collaborative filtering methods \cite{minto2021stronger,ammad2019federated} mainly focus on the cross-device FL setting. They adopt the idea of the typical federated learning algorithm FedAvg \cite{mcmahan2017communication} and average the updates of gradient information from participants. However, this line of methods has serious privacy issues. Sharing gradient information could lead to private data leakage as proofed by \cite{chai2020secure}. 

We notice that there is a contemporary work \cite{li2022semi} similar to our approach. The authors propose a cross-silo federated recommendation framework using split knowledge distillation. However, they assume that there are massive unlabeled overlapped data between parties and ignore privacy issues, which is a key concern in cross-silo FedRec. Exposing more overlapped data during training will inevitability lead to a higher risk of privacy leakage and largely increase the privacy budget.

\subsection{Federated Knowledge Distillation}

Knowledge distillation \cite{hinton2015distilling} distills knowledge from a teacher model to a student model through the soft target of the teacher model. More precisely, the student model learns from the teacher model by imitating the soft target distribution of the teacher model as model output through a Kullback-Leibler (KL) divergence loss defined as:
\begin{equation}
    L_{KD}(\boldsymbol{p},\boldsymbol{q}) = T^2 KL(\boldsymbol{p}||\boldsymbol{q}),
\end{equation}
\noindent where $\boldsymbol{p}$ and $\boldsymbol{q}$ are the soften output of the student model and teacher model. T is the temperature parameter. Denote the student logit as $\boldsymbol{z_s}$ and teacher logit as $\boldsymbol{z_t}$. Then $\boldsymbol{p}=softmax(\boldsymbol{z_s}/T)$ and $\boldsymbol{q}=softmax(\boldsymbol{z_t}/T)$.

Ensemble distillation \cite{you2017learning} ensembles knowledge from multiple teacher networks to yield better teacher knowledge. Current federated knowledge distillation methods \cite{li2019fedmd,lin2020ensemble,he2020group,gong2022preserving,li2020practical} adopts this idea to fuse heterogeneous local models into a global model and share model outputs instead of gradients. FedMD \cite{li2019fedmd} first trained student models through averaged logits of each sample on a public labeled dataset to reduce communication costs during training. Along this direction, Lin et al. \cite{lin2020ensemble} designed a more robust model distillation framework named FedDF, which allows for heterogeneous client models and data. 
Chang et al. \cite{he2020group} used group knowledge transfer to reduce the communication and computation overload for edge devices. Gong et al. \cite{gong2022preserving} proposed to use cross-domain unlabelled public data to protect private data privacy. These studies are designed for the cross-device setting and transfer knowledge through labeled public datasets \cite{li2019fedmd,ahn2020cooperative}, or unlabeled public datasets \cite{lin2020ensemble,gong2022preserving,li2020practical}. The improvement of their model performance unavoidably benefits from the extra public data. Li et al. \cite{li2020practical} proposed a practical one-shot federated learning algorithm for cross-silo FL setting. However, they still leverage an unlabeled public dataset to transfer knowledge.

Along with these studies, FedPDD is designed to transfer knowledge through limited overlapped user data for cross-silo FedRec. Unlike the above methods that only explore explicit knowledge from the other party, we propose a double distillation strategy that enables local models to fully exploit both explicit knowledge from the other party and implicit knowledge from itself.

\section{Methodology}\label{SectionApproach}

\subsection{Problem Statement}

Consider two parties A, B and a central server. Each party has a private labeled dataset $D^k := \{\boldsymbol{x_i^k}, y_i \}_{i=1}^{|D^k|}$ and can independently design its model $f^k$, $k = \{ A,B\}$. There exists limited overlapped users shared between two parties $D^c := \{\boldsymbol{x_i}^A, \boldsymbol{x_i}^B, y_i \}_{i=1}^{|D^c|}$. Let $\alpha = \frac{|D^c|}{|D^A|+|D^B|}$ denotes the overlapped data ratio. Our target is to improve the performance of the local models in both parties as well as the global model in the central server without exchanging raw data. Moreover, due to privacy concerns, the private information exposed to the server or the other party should be as less as possible. We summarize the notations used in this paper in table \ref{tab1}. 

\subsection{Overview of FedPDD}

The overall pipeline of FedPDD consists of three steps: pretraining, federated ensemble and local training. In the first stage, local models are pretrained on the local private datasets from scratch. In the federated ensemble stage, each party makes predictions on the overlapped user data using its pretrained local model. Then the central server aggregates the local predictions from two parties to obtain the ensemble teacher knowledge and distributes it to parties. The aggregation is protected by differential privacy as illustrated in section \ref{ensemble}. In the local training stage, we propose a double knowledge distillation strategy to fully explore both implicit knowledge from itself and explicit knowledge from the other party. That is to say, local models have two teachers and learn from multiple informative sources simultaneously during training. Therefore, by providing more teachers, our double distillation strategy enhances model performance and generalization ability. The details are illustrated in section \ref{selfdistill} and \ref{ensembledistill}.

We denote one local training and one federated ensemble as a round and repeat these two steps until models achieve convergence. Note that all training is performed offline and communications are only required in the federated ensemble stage. This offline learning strategy largely decreases communication needs and privacy leakage risk. Algorithm \ref{alg:algorithm} summarizes the whole training process of FedPDD.

\subsection{Distilling Implicit Knowledge\label{selfdistill}}
In order to fully explore the implicit knowledge of local models, we propose a self-distillation strategy to enable the local model to distill knowledge from itself. That is to say, the teacher is the student model itself. Specifically, the student in the previous rounds becomes the teacher of itself in the current round. The key idea is that we regard the previous student model outputs as different views of the features, which provides more information for the training \cite{allen2020towards}.

We design a simple but effective method to obtain the self-teacher model. We compare all intermediate local models in previous rounds as the candidates and select the one with the best performance as the teacher model for the current round. Considering deep learning models usually have a large model size, we only maintain one historical best model for each party and replace it whenever there is an improved one.

Denote the best local model on party $k$ in previous $n-1$ round as $f^k_{b_{(n-1)}}$. To learn from the implicit knowledge, we let the output of local model $f^k$ approximate the output of the teacher $f^k_{b_{(n-1)}}$ in round $n$ through the self-distillation (SD) loss $L_{SD}^k$. It is given by the KL divergence between local model output $\boldsymbol{p^k}$ and the teacher model output $\boldsymbol{p^k_{b}}$:
\begin{equation}
    \label{selfloss}
    \mathcal{L}_{SD}^k(\boldsymbol{p^k},\boldsymbol{p^k_b}) = T_{SD}^2 KL(\boldsymbol{p^k}||\boldsymbol{p^k_b}),
\end{equation}
\noindent where $T_{SD}$ is the self-distillation temperature.

\begin{algorithm}[tb]
\caption{Proposed FedPDD algorithm} 
\begin{algorithmic}[1]\label{alg:algorithm} 
\REQUIRE Local dataset $D_A$, $D_B$, overlapped dataset $D_c$, round number $n$, temperature $T$, trade-off weights $\beta$, $\gamma$, $w$.
\ENSURE Best local models $f^A_{b_{(n)}}$, $f^B_{b_{(n)}}$
\STATE Let $i=1$.
\WHILE{$i \leq n$}
\STATE // Perform local training
    \FOR{$k\in \{A, B\}$}
    \WHILE{not converge} 
        \STATE $f^k_{b_{(i)}}$ = $f^k_{b_{(i-1)}}$
	    \STATE Compute Loss based on equation 10
	    \STATE Compute gradients and update $f^k$
	    \IF{$f^k$ is better than $f^k_{b_{(i)}}$}
	    \STATE Update $f^k_{b_{(i)}} = f^k$
	    \ENDIF
	\ENDWHILE
	\ENDFOR
	\STATE // Perform federated ensemble
	\FOR{$k \in \{A,B\}$}
	    \FOR{each overlapped data $x_c$ in $D_c$}
	    \STATE Compute $\boldsymbol{z^k_c}$ and perturb with Gaussian noise
	    \STATE Send the perturbed logit $\boldsymbol{z'^k_c}$ to the server
	    \STATE Server computes the ensemble soft target distribution $\boldsymbol{p^t_c}$ and send it back to parties
	    \ENDFOR
	\ENDFOR
\ENDWHILE
\STATE \textbf{return} best local models $f^A_{b_{(n)}}$, $f^B_{b_{(n)}}$
\end{algorithmic}
\vspace{-1mm}
\end{algorithm}

\subsection{Distilling Explicit Knowledge\label{ensembledistill}}
We adopt ensemble distillation \cite{you2017learning} to leverage explicit knowledge from the other party through the overlapped user data. The key idea is that the ensemble of student models often yields improvements in system performance compared to the performance of individual models. To distill the explicit knowledge, we regard the ensemble results of local model predictions on the overlapped data as the ensemble teacher knowledge. By imitating this teacher knowledge, local models are able to learn from the other party.

Denote the ensemble teacher knowledge as $\boldsymbol{p^t_{c}}$. The ensemble distillation loss is given by the KL divergence between local model output $\boldsymbol{p^k}$ and the ensemble teacher knowledge:
\begin{equation}
    \label{ensembleloss}
    \mathcal{L}_{KD}^k(\boldsymbol{p^k},\boldsymbol{p^t_c}) = T_{ED}^2 KL(\boldsymbol{p^k}||\boldsymbol{p^t_c}),
\end{equation}
\noindent where $T_{ED}$ is the ensemble distillation temperature.

\subsection{Federated Ensemble}\label{ensemble}
During federated ensemble stage, communication is protected by two levels of privacy. First, we only send model outputs instead of sending model parameters or gradients. Second, if the central server is curious, directly updating local logits may have the risk of privacy leakage. Inspired by PATE \cite{papernot2016semi}, we perturb the local output logits with a Gaussian noise to ensure a higher privacy guarantee. 

Consider an overlapped sample $\boldsymbol{x_c} \in D^c$. In round $n$, party $k$ first presents local prediction through local best model $f^k_{b_{(n)}}$ obtained from local training. Denote the output logit of $f^k_{b_{(n)}}$ as $\boldsymbol{z_c^k}$:
\begin{equation}
    \boldsymbol{z_c^k} = f^k_{b_{(n)}}(\boldsymbol{x_c}).
\end{equation}
The perturbed ensemble logit $\boldsymbol{z_c'^t}$ is a linear combination of the perturbed local output logits of $f^k_{b_{(n-1)}}$. It can be expressed as:
\begin{gather}
 \label{equation5}
    \boldsymbol{z_c'^k} = f^k_{b_{(n)}}(\boldsymbol{x_c}) + \mathcal N(0,\sigma^2) \\
 \label{equation2}
    \boldsymbol{z_c'^t} = w \boldsymbol{z_c'^A} + (1-w)\boldsymbol{z_c'^B},
\end{gather}
where $w$ is the ensemble weight and $\sigma$ is the variance of Gaussian noise. Then the soft target distribution $\boldsymbol{p_c^t}$ of $\boldsymbol{x_c}$ can be defined as
\begin{gather}
    \label{equation3}
    \boldsymbol{p_c'^t} = \sigma_T(\boldsymbol{z_c'^t}) \\
    \label{equation4}
    \sigma_T(\boldsymbol{z_c'^t}) = \frac{\exp(\boldsymbol{z_c'^t}/T_{ED})}{\sum_{i=1}^n \exp(z_{ci}'^t/T_{ED})},
\end{gather}
\noindent where $\sigma_T$ is the general softmax function tuned by the ensemble temperature $T_{ED}$. The standard softmax function can be viewed as a special case of the general softmax function with $T=1$.

\subsection{Local Training}

We use an offline training scheme due to efficiency and privacy concerns. Offline training can largely reduce communication needs and therefore limit the exposure of private information to the server. During the local training, each party trains its local model $f^k$ parameterized by $f^k(\theta)$. $\theta^*$ is optimized by minimizing the training objective function $\mathcal{L}_{train}$:
\begin{equation}
    \theta^* = argmin_\theta \; \mathcal{L}_{train}.
\end{equation}

In this stage, we leverage three kinds of knowledge to enhance the model performance: direct knowledge from private labeled data, implicit knowledge from the best local model in the previous round and explicit knowledge from the other party. The direct knowledge is learned through the cross-entropy loss $\mathcal{L}_{CE}$ computed by the local model outputs $\boldsymbol{p^k}$ and the ground truth labels $y$. Our overall training objective function $\mathcal{L}_{train}$ is a weighted combination of three loss terms. Combining equation \ref{selfloss} and \ref{ensembleloss}, $\mathcal{L}_{train}$ can be written as:
\begin{equation}
	\mathcal{L}_{train} = \mathcal{L}_{SD} + \beta \mathcal{L}_{KD} + \gamma \mathcal{L}_{CE},
\end{equation}
\noindent where $\gamma$ and $\beta$ are the corresponding trade-off weights. We simply remain these weights to be unchanged during training. More experimental details are introduced in section \ref{setting}.

Note that local models have converged on their private data during the pretraining stage, which means that the initial best local models already contain valuable information. Therefore, both the self teacher knowledge and ensemble teacher knowledge are informative from the first round. 

\subsection{Inference}

In the inference phase, given a sample $x$, a party first checks whether the test sample $x$ is aligned with the other party. 
If the sample is aligned between both parties, two parties first infer locally through their obtained best local models and then ensemble the local predictions to give a joint prediction as the final result.
Otherwise, the party directly returns the prediction of its local model as the final result.

\subsection{Communication Analysis of FedPDD}
We analyze the communication cost of FedPDD in this section. Assume that the communication cost of updating or downloading a record (i.e. logit of the local prediction for $m$ class classification) from the server once is m. The total number of communication rounds until local models converge is n. Then the overall communication cost is $2mn|D^c|$. We can see that overall communication cost relates to three factors: the number of communication rounds, the amount of overlapped data involved in training and the size of updates. Our offline training strategy only has O(1) communication rounds as shown in section \ref{communication} and the size of model output $m$ is also O(1) for each record. Therefore, the communication cost of FedPDD is O($|D^c|$). In contrast, the online training strategy requires O(100) communication rounds and transmitting model information which usually involves thousands of parameters will cost more communication overload.

\subsection{Privacy Analysis of FedPDD}
In this section, we follow the prior works \cite{balle2018improving,dwork2006calibrating} and give the privacy analysis of FedPDD.

\begin{definition}(Differential Privacy \cite{dwork2006calibrating}). 
A randomized mechanism $M: \mathcal{X} \rightarrow \mathcal{Y}$ is $(\epsilon, \delta)$-DP if for every pair of datasets $X, X' \in \mathcal{X}$ that only differ in one sample, and every possible output $T\in \mathcal{Y}$. The following inequality holds:
\begin{equation}
\mathbb{P}[M(x) \in E] \leq e^{\varepsilon} \mathbb{P}\left[M\left(x^{\prime}\right) \in E\right]+\delta,
\end{equation}
where $\epsilon, \delta \geq 0$ are privacy loss parameters. 
\end{definition}

\begin{definition} ($l_2-$sensitivity). The $l_2$-sensitivity of a function $f: \mathcal{X} \rightarrow \mathbb{R}^d$ is
\begin{equation}
\Delta_{2}(f)=\max _{X, X^{\prime} \in \mathcal{X}}\left\|f(X)-f\left(X^{\prime}\right)\right\|_{2}.
\end{equation}
\end{definition}

\begin{definition}(Analytic Gaussian Mechanism~\cite{balle2018improving}). \label{DP}
Let $f:\mathcal{X} \rightarrow \mathbb{R}^d$ be a function with global $L_2$ sensitivity $\Delta$. For any $\epsilon \geq 0$ and $\delta \in [0,1]$, the Gaussian output perturbation mechanism $M(x)=f(x)+Z$ with $Z \sim \mathcal{N}(0,\sigma^2I)$ is $(\epsilon,\delta)-DP$ if and only if
\begin{equation}
\Phi(\frac{\Delta}{2\sigma}-\frac{\epsilon \sigma}{\Delta}) - e^{\epsilon}\Phi(-\frac{\Delta}{2\sigma}-\frac{\epsilon \sigma}{\Delta}) \leq \delta,
\end{equation}
where $\Phi$ is the CDF function of $\mathcal{N}(0,1)$.
\end{definition}

\begin{definition}(Composition of DP Algorithms~\cite{dwork2014algorithmic, dwork2009differential}).
Suppose $M=(M_1, M_2, ..., M_k)$ is a sequence of algorithms, where $M_i$ is $(\epsilon_i, \delta_i)$-DP, and the $M_i$'s are potentially chosen sequentially and adaptively. Then $M$ is $(\sum_{i=1}^{k} \epsilon, \sum_{i=1}^{k} \delta)$-DP. 
\end{definition}

For a meaningful privacy guarantee, we have $\delta = o(\frac{1}{n})$, where $n$ is the size of the dataset. By fixing privacy budget $\epsilon$, we calibrate the noise with proper $\sigma$ given by definition \ref{DP}. Therefore, our method preserves $(\epsilon,\delta)-DP$.

\begin{table*}[t]
\centering
  \caption{Main Result (overlapped data ratio $\alpha = 0.1$)}
  \label{main}
  \begin{tabular}{ccccccc}
    \toprule
     & \multicolumn{3}{c}{HetRec-MovieLens} & \multicolumn{3}{c}{Criteo} \\
    Model settings & Local A & Local B & Joint Prediction & Local A & Local B & Joint Prediction\\
    \midrule
    DeepFM~\cite{guo2017deepfm}       & 79.65\% & 80.36\% &    -   & 72.74\% & 72.93\% & - \\
    Ensemble~\cite{dietterich2002ensemble}              &    -   &    -   & 80.53\% & - & - & 73.08\% \\
    FTL~\cite{9076003} &    -   &    -   & 78.94\% & - & - & 73.90\% \\
    PFML~\cite{yang2021mutualpfl} & 80.26\% & 80.62\% &    81.32\%   & 74.28\% & 74.26\% & 75.53\% \\
    FedKD~\cite{gong2022preserving}         &  81.70\% & 81.60\% & 81.71\%  & 74.17\% &  74.28\% & 75.41\%  \\
    FedPDD (ours)        & \textbf{82.91\%} & \textbf{82.88\%} & \textbf{82.92\%} & \textbf{75.36\%} & \textbf{75.07\%} & \textbf{76.68\%} \\
    
  \bottomrule
\end{tabular}
\vspace{-2em}
\end{table*}

\begin{table}[t]
  \caption{Statistics of benchmark datasets}
  \label{tab2}
  
  \resizebox{\columnwidth}{!}{\begin{tabular}{ccccc}
    \toprule
    Name & \#users & \#movies & \#feat. fields & \#ratings\\
    \midrule
    HetRec-MovieLens  & 2,113 & 10,197 & 33 & 855,598\\
    Criteo  & - & - & 39 & 1000,000 \\
  \bottomrule
\end{tabular}
}
\vspace{-2em}
\end{table}

\section{Experiments}\label{experiment}
In this section, we evaluate our proposed FedPDD on two public real-world datasets. We aim to answer the following two questions through our experiments:
\begin{itemize}
    \item Q1: How well does FedPDD perform compared to the SOTA baselines on the benchmark datasets?
    \item Q2: How does differential privacy influence the performance of FedPDD?
\end{itemize}

\vspace{-2mm}
\subsection{Datasets}
The statistics of two benchmark public datasets are summarized in Table \ref{tab2}.

\textbf{HetRec-MovieLens Dataset} 
This dataset~\footnote{https://grouplens.org/datasets/hetrec-2011/} is a heterogeneous dataset for movie recommendation with 86W ratings. It is an extension of the MovieLens10M dataset, which contains personal ratings and tags about movies. The movies are linked to Internet Movie Database (IMDb) and RottenTomatoes movie review systems, greatly extending the feature space. The dataset is converted as a classification task by taking instances with a rating lower than 3 as negative instances, otherwise positive. In our setting, party A holds 21 features mainly coming from MovieLens, including user ID, movie ID, tags and etc., while party B holds 13 features from RottenTomatoes, including movie information, critics score, Rotten data and etc.

\textbf{Criteo} 
The Criteo dataset~\footnote{https://labs.criteo.com/category/dataset/} is generated from the original criteo dataset by randomly sampling 1 million instances. The task is to predict the ad click-through rate (CTR). It consists of 13 numerical features as well as 26 categorical features. We randomly split both numerical and categorical features into two parts with 19, 20 features respectively for two parties.

\subsection{Baselines}
We compare FedPDD with the following baselines:

\textbf{Locally trained DeepFM}:
DeepFM~\cite{guo2017deepfm} can handle high-order feature interaction of user embeddings and item embeddings in a centralized manner. Each party locally trains a DeepFM model on its own private dataset, and can not make use of the private features from the other party. 

\textbf{Ensemble of DeepFM}: 
Each party first locally trains a DeepFM model, then the local model outputs are aggregated to obtain the final joint prediction. We average the predictions from local models for model aggregation. 

\textbf{FTL baseline}: 
The FTL~\cite{9076003} approach maps the samples from heterogeneous feature spaces from two parties into a common latent space. Then, the two feature representations are concatenated and input to a classifier for prediction. We train a unique feature extractor for feature extraction in each party and then collaboratively train a classifier for label prediction. 



\textbf{PFML baseline}: 
The PFML~\cite{yang2021mutualpfl} approach integrates deep mutual learning~\cite{ying2018DML} into the local update process in each party to improve the performance of both the global model and the personalized local models.

\textbf{FedKD baseline}:
The FedKD~\cite{gong2022preserving} algorithm uses ensemble distillation for robust model fusion on heterogeneous local models. 
To transfer knowledge, an unlabeled dataset is used to sample data for all participants to compute logits and distill knowledge. We adopt this approach to our setting. In each communication round, all parties perform ensemble distillation, in which the local model parameters are evaluated on the aligned samples to generate logit outputs that are used to train each student model.

\vspace{-1mm}
\subsection{Experiment Settings}\label{setting}
For experiments, we randomly sample 80\% data as a training dataset and the rest for testing. Then we randomly split the dataset into two local datasets according to the overlapped data ratio. For the test dataset, we assume that all the data is aligned and shared by two parties.

We adopt DeepFM~\cite{guo2017deepfm} as the backbone, which is proposed to handle sophisticated feature interactions behind user behaviors for recommendation tasks. For training, we follow our two-stage training process as described in section~\ref{SectionApproach}. The model is optimized by Adam~\cite{kingma2014adam}. We set the learning rate for both local models to 0.001, the weight decay to 0.0001, the number of communication rounds to 5, and the batch size to 1024. The training process stops when the training achieves the maximum communication round. For experiments on the HetRec-MovieLens dataset, we set the temperature $T$ to 30, the trade-off loss weights $\beta$, $\gamma$ to 10, and the ensemble weight $w$ to 0.5. For experiments on the Criteo dataset, we set the temperature $T$ to 30, the trade-off loss weights $\beta$, $\gamma$ to 3, and the ensemble weight $w$ to 0.5. For all the ablation settings, we conduct experiments three times and report the average.

We use accuracy as the metric to evaluate our experiment results. The closer the value of accuracy to 1, the better the performance of prediction is.

\subsection{Main Results}

\textbf{Comparison with local training}.
To demonstrate the effectiveness of our proposed method, we first compare it with the local fine-tuned baselines (i.e. Local A and Local B in Table \ref{main}). From Table \ref{main}, it is observed that 1) both local models benefit significantly from our approach, with an increment of 3.26\%/2.44\% on the HetRec-MovieLens and 2.62\%/2.14\% on Criteo. 2) The joint prediction further brings extra performance gain to the local models by around 0.04\% and 1.61\%, respectively. These results show that our approach successfully transfers knowledge between two parties and therefore improves the performance of their local models.

\begin{figure}[t]
\centering
\subfigure[HetRec-MovieLens]{
\includegraphics[width=0.46\columnwidth]{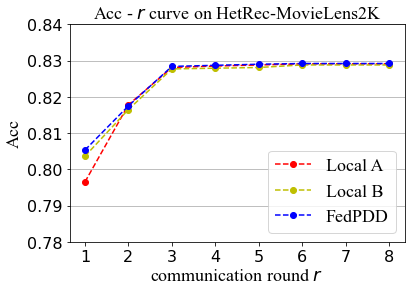}
}
\subfigure[Criteo]{
\includegraphics[width=0.46\columnwidth]{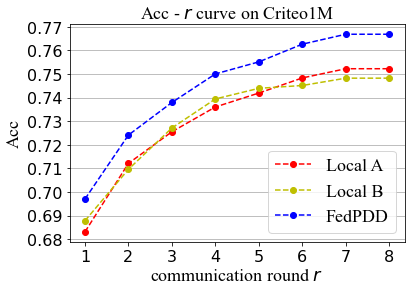}
}

\caption{The relationship between communication round $r$ and performance of FedPDD during training}
\label{figure2}
\vspace{-1em}
\end{figure}

\begin{figure}[t]
\centering
\subfigure[HetRec-MovieLens]{
\includegraphics[width=0.46\columnwidth]{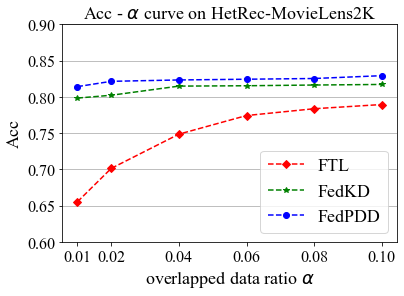}
}
\subfigure[Criteo]{
\includegraphics[width=0.46\columnwidth]{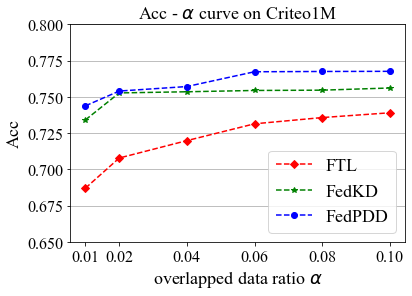}
}
\caption{The comparison between FedPDD and FTL baseline when the overlapped data ratio $\alpha$ decreases}
\label{figure3}
\vspace{-2em}
\end{figure}

\textbf{Comparison with SOTAs}.
FTL-based methods \cite{9076003} leverage the overlapped data to make a joint prediction. Therefore, we only compare it with the joint prediction of FedPDD. From Table \ref{main}, we can see that the joint prediction of FedPDD is better than FTL by 3.98\% on HetRec-MovieLens and 2.78\% on Criteo. These results show that our proposed framework outperforms the FTL-based methods in the situation where overlapped data is limited.

We also compare FedPDD with two knowledge distillation-based federated learning strategies PFML \cite{yang2021mutualpfl} and FedKD~\cite{gong2022preserving}. From Table \ref{main}, we can find that the local model performance of FedPDD outperforms the PFML baseline by 2.46\% and 1.92\% on two datasets on average. Besides, our FedPDD outperforms FedKD by an additional 1.25\% and 0.99\% on two datasets on average. This indicates that our double distillation method can generate better teacher logits from not only the ensemble of cross-party local models but also the previous local models of the same party, thereby effectively enhancing the local model performance.

For joint prediction on aligned test samples, the predictions of both local models are averaged as the final result. Therefore, the performance of federated joint prediction mainly depends on the performance of local models. From Table~\ref{main}, we can observe that FedPDD outperforms FedKD by 1.21\% and 1.27\% on two datasets, respectively, and outperforms PFML by 1.60\% and 1.15\% on two datasets, respectively. This is reasonable as the local models trained by FedPDD achieve higher accuracy than FedKD and PFML on both datasets. Meanwhile, more accurate joint predictions can in turn transfer more knowledge to both local models.

\begin{figure}[t]
\centering
\includegraphics[width=0.7\columnwidth]{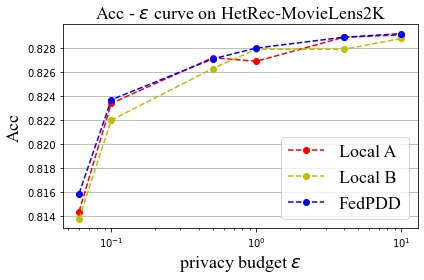}
\caption{The impact of DP parameter $\epsilon$ on model performance on HetRec-MovieLens dataset.}
\label{figureDP}
\vspace{-1em}
\end{figure}

\begin{table}[t]
\footnotesize
\setlength{\tabcolsep}{1.5mm}
\centering
  \caption{Sensitivity to $T$ on HeterRec-MovieLens2K}
  \label{tab3}
  \begin{tabular}{ccccccc}
    \toprule
    $T$ & 1 & 10 & 20 & 30 & 40 & 50 \\
    \midrule
    FedPDD    & 0.8258 & 0.8260 & 0.8270 & \textbf{0.8291} & 0.8270 & 0.8267 \\
    Local A  & 0.8254 & 0.8252 & 0.8260 & \textbf{0.8288} & 0.8259 & 0.8256 \\
    Local B  & 0.8268 & 0.8261 & 0.8270 & \textbf{0.8292} & 0.8270 & 0.8267 \\
  \bottomrule
\end{tabular}
\vspace{-2em}
\end{table}

\begin{table}[t]
\footnotesize
\setlength{\tabcolsep}{1.5mm}
\centering
  \caption{Sensitivity to $T$ on Criteo}
  \label{tab4}
  \begin{tabular}{ccccccc}
    \toprule
    $T$ & 1 & 10 & 20 & 30 & 40 & 50 \\
    \midrule
    FedPDD    & 0.7520 & 0.7508 & 0.7521 & \textbf{0.7522} & 0.7514 & 0.7507 \\
    Local A  & 0.7473 & 0.7487 & \textbf{0.7489} & 0.7482 & 0.7481 & 0.7480 \\
    Local B  & 0.7657 & 0.7664 & 0.7664 & \textbf{0.7668} & 0.7660 & 0.7663 \\
  \bottomrule
\end{tabular}
\vspace{-2em}
\end{table}

\subsection{Hyper-parameter Tuning}

\subsubsection{Effect of communication round $r$}\label{communication}
From Figure \ref{figure2}, we can find that FedPDD converges after 5 rounds on the HetRec-MovieLens dataset and 7 rounds on the Criteo dataset, which demonstrates that FedPDD only requires a few times of communication need between two parties during training.

\subsubsection{Effect of overlapped data ratio $\alpha$}
As we mentioned previously, the major challenge of the multi-view federated learning problem is that the overlapped data is often limited. We adjust the overlapped data ratio $\alpha$ from $0.1$ to $0.01$ to test the effectiveness of FedPDD. The experimental results are shown in Figure \ref{figure3}. We can observe that when $\alpha$ decreases to 0.01, the performance of the FTL approach drops significantly by 13.42\% while FedPDD still remains above 0.8 on the HetRec-MovieLens dataset. On the Criteo dataset, FedPDD almost remains the same while the FTL baseline drops around 5.20\% when $\alpha$ decreases. These results demonstrate the effectiveness of FedPDD in our setting.

\subsubsection{Effect of temperature $T$}
In knowledge distillation, the temperature is used to soften the probability output, leading the students to pay more attention to the small number of logits \cite{hinton2015distilling}. In this section, we conduct experiments to find out the influence of temperature on our model. We let $T = T_{SD} = T_{ED}$. In Table \ref{main}, we set the temperature to 30 to highlight the best performance of FedPDD on two benchmark datasets. Here we fine-tune the temperature $T$ over a large range from 1 to 50 in the online scheme. The results in Table \ref{tab3} and \ref{tab4} show that the careful selection of temperature can bring a little performance enhancement on local models.

\subsubsection{Effect of differential privacy budget $\epsilon$} In Figure 5, we demonstrate the effect of differential privacy on the HetRec-MovieLens dataset. We change the $\epsilon$ from 0.05 to 10 to explore the change in local model performance and federated model performance. 
It can be observed that the accuracy drops only by 1.53\% for local models in FedPDD and 1.30\% for joint prediction of FedPDD, respectively. The performance of local models outperforms locally trained baselines when $\epsilon > 0.05$.

\section{Conclusion}\label{conclusion}
In this paper, we propose a novel cross-silo federated recommendation framework FedPDD. We design a double distillation strategy that leverages knowledge not only from the ensemble of local models but also from previous local models to efficiently improve the model performance. Besides, FedPDD largely reduces communication needs and privacy leakage risk by utilizing an offline training strategy and only transmitting model output during training. Additionally, differential privacy is introduced to protect the communication process with a higher level of privacy protection. 
Experimental results demonstrate that our approach can effectively exploit both implicit knowledge and explicit knowledge and thereby enhance the performances of both local and overall joint prediction tasks. Moreover, our framework can also be adopted to learn and predict financial risks associated with various internet finance platforms with heterogeneous information features and strong privacy-preserving needs.

\section{Acknowledgment}
The authors gratefully acknowledge funding from Guangdong Province Focus Research Project (Grant Number: 2019KZDZX2014), Guangdong Province Research Fund (Grant Number: 2019QN01X277), National Natural Science Foundation of China (Grant Numbers: 71971106, 72001099), and Shenzhen Humanities \& Social Sciences Key Research Bases. We would like to show our gratitude to Guangneng Hu, Ce Ju, Ben Tan and Prof. Qiang Yang for their advice on the earlier manuscript and we thank all the reviewers for valuable comments.

\bibliographystyle{IEEEtran}
\bibliography{sample-base}

\begin{thebibliography}{10}
\providecommand{\url}[1]{#1}
\csname url@samestyle\endcsname
\providecommand{\newblock}{\relax}
\providecommand{\bibinfo}[2]{#2}
\providecommand{\BIBentrySTDinterwordspacing}{\spaceskip=0pt\relax}
\providecommand{\BIBentryALTinterwordstretchfactor}{4}
\providecommand{\BIBentryALTinterwordspacing}{\spaceskip=\fontdimen2\font plus
\BIBentryALTinterwordstretchfactor\fontdimen3\font minus
  \fontdimen4\font\relax}
\providecommand{\BIBforeignlanguage}[2]{{%
\expandafter\ifx\csname l@#1\endcsname\relax
\typeout{** WARNING: IEEEtran.bst: No hyphenation pattern has been}%
\typeout{** loaded for the language `#1'. Using the pattern for}%
\typeout{** the default language instead.}%
\else
\language=\csname l@#1\endcsname
\fi
#2}}
\providecommand{\BIBdecl}{\relax}
\BIBdecl

\bibitem{zhang2019deep}
S.~Zhang, L.~Yao, A.~Sun, and Y.~Tay, ``Deep learning based recommender system:
  A survey and new perspectives,'' \emph{ACM Computing Surveys (CSUR)},
  vol.~52, no.~1, pp. 1--38, 2019.

\bibitem{yang2019federated}
Q.~Yang, Y.~Liu, T.~Chen, and Y.~Tong, ``Federated machine learning: Concept
  and applications,'' \emph{ACM Transactions on Intelligent Systems and
  Technology (TIST)}, vol.~10, no.~2, pp. 1--19, 2019.

\bibitem{kairouz2019advances}
P.~Kairouz, H.~B. McMahan, B.~Avent, A.~Bellet, M.~Bennis, A.~N. Bhagoji,
  K.~Bonawitz, Z.~Charles \emph{et~al.}, ``Advances and open problems in
  federated learning,'' \emph{arXiv preprint arXiv:1912.04977}, 2019.

\bibitem{9076003}
Y.~Liu, Y.~Kang, C.~Xing, T.~Chen, and Q.~Yang, ``A secure federated transfer
  learning framework,'' \emph{IEEE Intelligent Systems}, vol.~35, no.~4, pp.
  70--82, 2020.

\bibitem{Siwei2020mmvfl}
F.~Siwei and Y.~Han, ``Multi-participant multi-class vertical federated
  learning,'' \emph{arXiv preprint arXiv:2001.11154}, 2020.

\bibitem{kang2020fedmvt}
Y.~Kang, Y.~Liu, and T.~Chen, ``Fedmvt: Semi-supervised vertical federated
  learning with multiview training,'' \emph{arXiv preprint arXiv:2008.10838},
  2020.

\bibitem{zhu2019deep}
L.~Zhu, Z.~Liu, and S.~Han, ``Deep leakage from gradients,'' in \emph{NeurIPS},
  vol.~32, 2019.

\bibitem{geiping2020inverting}
J.~Geiping, H.~Bauermeister, H.~Dr{\"o}ge, and M.~Moeller, ``Inverting
  gradients-how easy is it to break privacy in federated learning?'' in
  \emph{NeurIPS}, vol.~33, 2020, pp. 16\,937--16\,947.

\bibitem{sun2021soteria}
J.~Sun, A.~Li, B.~Wang, H.~Yang, H.~Li, and Y.~Chen, ``Soteria: Provable
  defense against privacy leakage in federated learning from representation
  perspective,'' in \emph{CVPR}, 2021, pp. 9311--9319.

\bibitem{abadi2016deep}
M.~Abadi, A.~Chu, I.~Goodfellow, H.~B. McMahan, I.~Mironov, K.~Talwar, and
  L.~Zhang, ``Deep learning with differential privacy,'' in \emph{Proceedings
  of the 2016 ACM SIGSAC Conference on Computer and Communications Security},
  2016, pp. 308--318.

\bibitem{mcmahan2017communication}
B.~McMahan, E.~Moore, D.~Ramage, S.~Hampson, and B.~A. y~Arcas,
  ``Communication-efficient learning of deep networks from decentralized
  data,'' in \emph{Artificial intelligence and statistics}.\hskip 1em plus
  0.5em minus 0.4em\relax PMLR, 2017, pp. 1273--1282.

\bibitem{ying2020shared}
S.~Ying, ``Shared mf: A privacy-preserving recommendation system,'' \emph{arXiv
  preprint arXiv:2008.07759}, 2020.

\bibitem{yang2021practical}
L.~Yang, B.~Tan, B.~Liu, V.~W. Zheng, K.~Chen, and Q.~Yang, ``Practical and
  secure federated recommendation with personalized masks,'' \emph{arXiv
  preprint arXiv:2109.02464}, 2021.

\bibitem{chai2020secure}
D.~Chai, L.~Wang, K.~Chen, and Q.~Yang, ``Secure federated matrix
  factorization,'' \emph{IEEE Intelligent Systems}, vol.~36, no.~5, pp. 11--20,
  2020.

\bibitem{yang2021fcmf}
E.~Yang, Y.~Huang, F.~Liang, W.~Pan, and Z.~Ming, ``Fcmf: Federated collective
  matrix factorization for heterogeneous collaborative filtering,''
  \emph{Knowledge-Based Systems}, vol. 220, p. 106946, 2021.

\bibitem{eren2022fedsplit}
M.~E. Eren, L.~E. Richards, M.~Bhattarai, R.~Yus, C.~Nicholas, and B.~S.
  Alexandrov, ``Fedsplit: One-shot federated recommendation system based on
  non-negative joint matrix factorization and knowledge distillation,''
  \emph{arXiv preprint arXiv:2205.02359}, 2022.

\bibitem{minto2021stronger}
L.~Minto, M.~Haller, B.~Livshits, and H.~Haddadi, ``Stronger privacy for
  federated collaborative filtering with implicit feedback,'' in
  \emph{Fifteenth ACM Conference on Recommender Systems}, 2021, pp. 342--350.

\bibitem{ammad2019federated}
M.~Ammad-Ud-Din, E.~Ivannikova, S.~A. Khan, W.~Oyomno, Q.~Fu, K.~E. Tan, and
  A.~Flanagan, ``Federated collaborative filtering for privacy-preserving
  personalized recommendation system,'' \emph{arXiv preprint arXiv:1901.09888},
  2019.

\bibitem{li2022semi}
W.~Li, Q.~Xia, J.~Deng, H.~Cheng, J.~Liu, K.~Xue, Y.~Cheng, and S.-T. Xia,
  ``Semi-supervised cross-silo advertising with partial knowledge transfer,''
  \emph{arXiv preprint arXiv:2205.15987}, 2022.

\bibitem{hinton2015distilling}
G.~Hinton, O.~Vinyals, and J.~Dean, ``Distilling the knowledge in a neural
  network,'' \emph{arXiv preprint arXiv:1503.02531}, 2015.

\bibitem{you2017learning}
S.~You, C.~Xu, C.~Xu, and D.~Tao, ``Learning from multiple teacher networks,''
  in \emph{Proceedings of the 23rd ACM SIGKDD}, 2017, pp. 1285--1294.

\bibitem{li2019fedmd}
D.~Li and J.~Wang, ``Fedmd: Heterogenous federated learning via model
  distillation,'' \emph{arXiv preprint arXiv:1910.03581}, 2019.

\bibitem{lin2020ensemble}
T.~Lin, L.~Kong, S.~U. Stich, and M.~Jaggi, ``Ensemble distillation for robust
  model fusion in federated learning,'' \emph{NeurIPS}, vol.~33, pp.
  2351--2363, 2020.

\bibitem{he2020group}
C.~He, M.~Annavaram, and S.~Avestimehr, ``Group knowledge transfer: Federated
  learning of large cnns at the edge,'' \emph{NeurIPS}, vol.~33, pp.
  14\,068--14\,080, 2020.

\bibitem{gong2022preserving}
X.~Gong, A.~Sharma, S.~Karanam, Z.~Wu, T.~Chen, D.~Doermann, and A.~Innanje,
  ``Preserving privacy in federated learning with ensemble cross-domain
  knowledge distillation,'' in \emph{AAAI}, vol.~36, no.~11, 2022, pp.
  11\,891--11\,899.

\bibitem{li2020practical}
Q.~Li, B.~He, and D.~Song, ``Practical one-shot federated learning for
  cross-silo setting,'' \emph{arXiv preprint arXiv:2010.01017}, vol.~1, no.~3,
  2020.

\bibitem{ahn2020cooperative}
J.-H. Ahn, O.~Simeone, and J.~Kang, ``Cooperative learning via federated
  distillation over fading channels,'' in \emph{IEEE ICASSP}.\hskip 1em plus
  0.5em minus 0.4em\relax IEEE, 2020, pp. 8856--8860.

\bibitem{allen2020towards}
Z.~Allen-Zhu and Y.~Li, ``Towards understanding ensemble, knowledge
  distillation and self-distillation in deep learning,'' \emph{arXiv preprint
  arXiv:2012.09816}, 2020.

\bibitem{papernot2016semi}
N.~Papernot, M.~Abadi, U.~Erlingsson, I.~Goodfellow, and K.~Talwar,
  ``Semi-supervised knowledge transfer for deep learning from private training
  data,'' \emph{arXiv preprint arXiv:1610.05755}, 2016.

\bibitem{balle2018improving}
B.~Balle and Y.~Wang, ``Improving the gaussian mechanism for differential
  privacy: Analytical calibration and optimal denoising,'' in
  \emph{ICML}.\hskip 1em plus 0.5em minus 0.4em\relax PMLR, 2018, pp. 394--403.

\bibitem{dwork2006calibrating}
C.~Dwork, F.~McSherry, K.~Nissim, and A.~Smith, ``Calibrating noise to
  sensitivity in private data analysis,'' in \emph{Theory of cryptography
  conference}.\hskip 1em plus 0.5em minus 0.4em\relax Springer, 2006, pp.
  265--284.

\bibitem{dwork2014algorithmic}
C.~Dwork, A.~Roth \emph{et~al.}, ``The algorithmic foundations of differential
  privacy,'' \emph{Foundations and Trends{\textregistered} in Theoretical
  Computer Science}, vol.~9, no. 3--4, pp. 211--407, 2014.

\bibitem{dwork2009differential}
C.~Dwork and J.~Lei, ``Differential privacy and robust statistics,'' in
  \emph{Proceedings of the forty-first annual ACM symposium on Theory of
  computing}, 2009, pp. 371--380.

\bibitem{guo2017deepfm}
H.~Guo, R.~Tang, Y.~Ye, Z.~Li, and X.~He, ``Deepfm: a factorization-machine
  based neural network for ctr prediction,'' \emph{arXiv preprint
  arXiv:1703.04247}, 2017.

\bibitem{dietterich2002ensemble}
T.~G. Dietterich \emph{et~al.}, ``Ensemble learning,'' \emph{The handbook of
  brain theory and neural networks}, vol.~2, no.~1, pp. 110--125, 2002.

\bibitem{yang2021mutualpfl}
R.~Yang, J.~Tian, and Y.~Zhang, ``Regularized mutual learning for personalized
  federated learning,'' in \emph{Proceedings of The 13th Asian Conference on
  Machine Learning}, vol. 157.\hskip 1em plus 0.5em minus 0.4em\relax PMLR,
  17--19 Nov 2021, pp. 1521--1536.

\bibitem{ying2018DML}
Y.~Zhang, T.~Xiang, T.~M. Hospedales, and H.~Lu, ``Deep mutual learning,'' in
  \emph{CVPR}, 2018.

\bibitem{kingma2014adam}
D.~P. Kingma and J.~Ba, ``Adam: A method for stochastic optimization,''
  \emph{arXiv preprint arXiv:1412.6980}, 2014.

\end{thebibliography}

\end{document}